\documentstyle[12pt]{article}
\begin{document}
\vskip 2 cm
\begin{center}
\Large{\bf ELECTRIC CHARGE, EARLY UNIVERSE AND THE 
SUPERSTRING THEORIES }
\end{center}
\vskip 3 cm
\begin{center}
{\bf AFSAR ABBAS} \\
Institute of Physics, Bhubaneswar-751005, India \\
(e-mail : afsar@iopb.res.in)
\end{center}
\vskip 20 mm  
\begin{centerline}
{\bf Abstract }
\end{centerline}
\vskip 3 mm

Very recently, it has been shown by the author that the Standard Model
Higgs cannot be a physical particle. Here, on most general grounds it is
established that as per the Standard Model there is no electric
charge above the electro-weak phase transition temperature. Hence there
was no electric charge present in the early universe. The Superstring
Theories are flawed in as much as they are incompatible with this
requirement. Hence the Superstring Theories are inconsistent with this
basic structure and requirement of the Standard Model.

\newpage

Contrary to earlier expectations, it has recently been demonstrated that
electric charge is quantized in the Standard Model (SM) [ 1,2 ]. This
quantization requires complete machinery of the SM including the Higgs. If
one or more conditions are not properly taken care of it may lead to
problems [ 2 ]. For example, analyses of the property of the electric
charge quantization in the SM, have led some authors to conclude that
millicharged particles are permitted by the SM. This is erroneous and
was pointed out recently by the author [ 3 ].

Besides other built-in features of the SM that go into the
demonstration of the electric charge quantization,
one crucial feature was spontaneous symmetry breaking through a
Higgs doublet [ 1,2 ]. Using this property, it was shown that when the
electro-weak symmetry is restored there is no electric charge [ 4 ]. Hence
there was no electric charge in the early universe. In this demonstration
one needs a Higgs doublet, as is present in the SM.

 Very recently [ 5 ], the author has shown that the SM Higgs is not a
physical particle. It is a manifestation of the `vacuum' which gives the
basic structure to the SM. In this demonstration [ 5 ] the Higgs was not a
doublet but had unconstrained isospin and hypercharge representations
(see [ 5 ] for details ). Here we ask the question, with this generalized
picture what happens to the electric charge when the symmetry is restored.
How does this result reflect upon the well studied and the
ever-promising Superstring Theories ?

It has been shown by the author [ 5 ] that in SM 
$ SU(N_{C}) \otimes SU(2)_{L} \times U(1)_{Y} $
for $ N_{C} = 3 $
spontaneous symmetry breaking by a Higgs of weak hypercharge $ Y_{ \phi } $
and general isospin T where $ T_{ 3 }^{ \phi } $ component 
develops the vacuum
expectation value $ < \phi >_{0} $, fixes `h' in the electric charge
definition $ Q = T_{ 3 } + h Y $ to give

\begin{equation}
Q = T_{ 3 } - \frac{ T_{ 3 }^{ \phi } }{ Y_{ \phi } } \, Y
\end{equation}

where Y is the hypercharge for doublets and singlets for
a single generation.
For each generation renormalizability through triangular anomaly
cancellation and the requirement of the identity of L- and R-handed charges
in $ U(1)_{ em } $ one finds that all unknown hypercharges are
proportional to 
$ \frac{ Y_{\phi} }{ T_{ 3 }^{ \phi } } $. Hence correct charges (for 
$ N_{C} = 3 $ ) fall through as below

\begin{eqnarray}
Q(u) = \frac{1}{2} ( 1 + \frac{1}{N_{C}} ) \nonumber \\
Q(d) = \frac{1}{2} ( -1 + \frac{1}{N_{C}} ) \nonumber \\
Q(e) = -1 \nonumber \\
Q(\nu) = 0
\end{eqnarray}

In this demonstration of charge quantization the isospin and the
hypercharge of Higgs were left completely unconstrained [ 5 ]. It was then
shown that the complete structure of the Standard Model was reproduced
without specifying any quantum numbers of the Higgs. Thus Higgs is unlike
any particle known to us. Hence it was predicted [ 5 ] by the author that
the Higgs is not a physical particle.

Note that the expression for Q in (1) arose due to spontaneous symmetry
breaking of $ SU(N_{C}) \otimes SU(2)_{L} \times U(1)_{Y} $
(for $ N_{C} = 3 $ ) to 
$ SU(N_{C}) \times U(1)_{em} $ through the medium of a Higgs with
arbitrary isospin T and hypercharge $ Y_{\phi} $. What happens when at
higher temperature, as for example found in the early universe, the 
$ SU(N_{C}) \otimes SU(2)_{L} \otimes U(1)_{Y} $ symmetry is restored ?
Then the parameter `h' in the electric charge definition remains
undetermined. Note that `h' was fixed as in (1) due to spontaneous
symmetry breaking through Higgs. Without it `h' remains unknown. 

 Earlier [ 4 ] we had found this for a Higgs doublet. Now we find that for
a Higgs with any arbitrary isospin and hypercharge, this is still true. As
`h' is not defined, the electric charge is not defined. Hence when the
electroweak symmetry is restored, irrespective of the Higgs isospin and
hypercharge the electric charge
disappears as a physical quantity. Hence here too we find that there was no
electric charge in the early universe.

In fact the property that there was no electric charge in the early
universe can be understood this way. The fact that L-charges were equal
to the R-charges is a property of $ U(1)_{em} $. This property need not
hold if $ U(1)_{em}$ is not present. And indeed when the symmetry is
restored to 
$ SU(2)_{L} \otimes U(1)_{Y} $ (at higher temperatures and/or early
universe) the L-charges and the R-charges (if they exist at all) need not
be equal to each other. If different, how different ? These lead to
self-contradictions. These contradictions can be prevented only if
there were no charges present when 
$ U(1)_{em} $ was not present. And this is what has been consistently
demonstrated here on most general grounds.

Let us now look at the Superstring Theories. What is the structure of
electric charge in Superstring Theories and how does it relate to the
Standard Model discussed above and in addition to Refs [ 1-5 ].

 The appearances of color neutral string states carrying fractional
electric charge (eg. $ q = \pm \frac{1}{2}, \pm \frac{1}{3} $) is a well
known problem in string theory [ 6,7 ]. This appears to be generic in
string theories. In string theory, in fact, there are always fractionally
charged states unless the group `levels' obey the condition [ 8 ] 

\begin{equation}
3 k_{1} + 3 k_{2} + 4 k_{3} = 0 mod 12
\end{equation} 

This condition is indeed obeyed by the standard GUT choice 
$ k_{1} = \frac{5}{3}, k_{2} = k_{3} = 1 $. However, at low energy one
would require the SM group 
$ SU(3) \otimes SU(2) \otimes U(1) $
to survive. Here this does not happen. An exact $ SU(5) $ symmetry
survives down to low energies and this is unwelcome. Thus, one has the
following possibilities
(1) One accepts the canonical possible values of the $ k_{i} $'s and try
to justify the existence of fractionally charged states at low energies.
This is in conflict with some experimental constraints on terrestrial
fractional charges. (2) One tries to choose non-canonical $ k_{i} $'s.
This idea then runs into conflict with other established ideas of
extensions beyond the SM.
(3) One chooses the canonical value for the $ k_{i} $'s and fractionally
charged particles are shunted to very high mass states. Hence present
experiments cannot touch them.

 All these are well known problems of the electric charge in the
Superstring Theories [ 9 ]. However here attention is drawn to the fact
that all putative extensions of the Standard Model should reduce smoothly
and consistently to the Standard Model at low energies. Not only that, all
these extensions should be consistent with the predictions of the Standard
Model at very high temperatures. Contrary to naive expectations, the SM
does make specific predictions at very high temperatures too.
For example one clear-cut prediction of the Standard Model as shown here
and also shown earlier [ 4 ], is that
at high enough temperatures (as in the early universe) when the unbroken 
$ SU(3) \otimes SU(2) \otimes U(1) $ symmetry is restored, there is no
electric charge. GUTs and other standard extensions of the SM are
incompatible with this requirement [ 3,4 ].

What about Superstring Theories ? Quite clearly generically in
Superstring Theories electric charge exists right up to the Planck Scale.
Hence as per this theory the electric charge, as an inherent property of
matter, has existed right from the beginning.
This is not correct in the SM. As shown here and earlier [ 4 ]
the electric charge came into existence at a later stage in the evolution
of the Universe when the  $ SU(2)_{L} \otimes U(1)_{Y} $
group was spontaneously broken to $ U(1)_{em} $. 
It was never there all the time. This is because electric charge is a
derived quantity [ 4 ]. Hence we find that in this regard the Superstring
Theories are inconsistent with the SM.

We have shown that the structure of the electric charge and its
quantization property in the Standard Model are very restrictive. It turns
out that even more demanding and restrictive is the property that there
was no electric charge before the spontaneous symmetry breaking of the
electro-weak group. Any theory which is
incompatible with this requirement cannot be a valid theory of nature.
Along with most (all ?) extensions of the SM [ 3,4 ], Superstring Theories
also fall in this category. In spite of their promise, the Superstring
Theories have this fatal flaw built into them.

\newpage

\begin{center}
{\bf\large REFERENCES }
\end{center}

\vskip 2 cm

1. A. Abbas, 
{\it Phys. Lett. } {\bf B 238} (1990) 344

2. A. Abbas,
{\it J. Phys. G. } {\bf 16 } (1990) L163

3. A. Abbas,
{\it Physics Today }, July 1999, p.81-82

4. A. Abbas, {\it `Phase transition in the early universe and charge
quantization'}; hep-ph/9503496

5. A. Abbas, {\it `What is the Standard Model Higgs ?'} ; 
hep-ph/9912243

6. X. G. Wen and E. Witten,
{\it Nucl. Phys. } {\bf B 261} (1985) 651

7. G. Athanasiu, J.Atick, M.Dine and W. Fischler, 
{\it Phys. Letts.} {\bf B 214} (1988) 55

8. A. N. Schellekens, {\it Phys. Letts.}
{\bf B 237} (1990) 363

9. J. Polchinski, {\it `String Theory'}, Cambridge University Press, 1998

\end{document}